\documentclass[twocolumn,final]{svjour3}

\smartqed

\usepackage{graphicx,amssymb}
\usepackage{epsf,psfig}
\usepackage{txfonts}
\usepackage{natbib}

\journalname{Nonlinear Dynamics}

\begin{document}

\title{Order and chaos in a new 3D dynamical model describing motion in non axially symmetric galaxies}

\author{Euaggelos E. Zotos \and Nicolaos D. Caranicolas}

\institute{Department of Physics, \\
Section of Astrophysics, Astronomy \& Mechanics, \\
Aristotle University of Thessaloniki \\
GR-541 24, Thessaloniki, Greece \\
Corresponding author's email: {evzotos@physics.auth.gr}
}

\date{Received: 25 June 2013 / Accepted: 1 August 2013 / Published online: 7 September 2013}

\titlerunning{Order and chaos in a new 3D dynamical model describing motion in non axially symmetric galaxies}

\authorrunning{E. E. Zotos \& N. D. Caranicolas}

\maketitle

\begin{abstract}

We present a new dynamical model describing 3D motion in non axially symmetric galaxies. The model covers a wide range of galaxies from a disk system to an elliptical galaxy by suitably choosing the dynamical parameters. We study the regular and chaotic character of orbits in the model and try to connect the degree of chaos with the parameter describing the deviation of the system from axial symmetry. In order to obtain this, we use the Smaller ALingment Index (SALI) technique by numerically integrating the basic equations of motion, as well as the variational equations for extensive samples of orbits. Our results suggest, that the influence of the deviation parameter on the portion of chaotic orbits strongly depends on the vertical distance $z$ from the galactic plane of the orbits. Using different sets of initial conditions, we show that the chaotic motion is dominant in galaxy models with low values of $z$, while in the case of stars with large values of $z$ the regular motion is more abundant, both in elliptical and disk galaxy models.

\keywords{Galaxies: kinematics and dynamics; New models}

\end{abstract}

\section{Introduction}
\label{intro}

To study the dynamical behavior of galaxies it is necessary to build a model describing the properties of the system. Information on the construction of dynamical models of galaxies are provided usually from observations, made after some necessary simplifying assumptions. On the other hand, it is a fact that galactic dynamical models are successful and realistic, only if the modeled characteristics agree with the corresponding observational data.

In most cases, the galactic models are spherical or axially symmetric, in an attempt to simplify the study of orbits. For instance, in a spherical system all three components of the angular momentum and of course the total angular momentum is conserved. Thus, we have a plane motion, which takes place in the plane perpendicular to the vector of the total angular momentum. On the other hand, in an axially symmetric system, where the motion is described by a potential $\Phi(R,z)$ the $L_{\rm z}$ component of the angular momentum is conserved and the motion takes place in the meridional plane $(R,z)$, which rotates non-uniformly around the axis of symmetry with angular velocity $\dot{\phi} = L_{\rm z}/R^2$.

Spherical models for galaxies were studied by [\citealp{12},\citealp{22},\citealp{33}]. Moreover, interesting axially symmetric galaxy models were presented and studied by [\citealp{11}]. Recently, [\citealp{34}] used data derived from rotation curves of real galaxies, in order to construct a new axially symmetric model describing the motion in both elliptical and disk galaxy systems.

Of particular interest, are the so-called composite galactic dynamical models. In those models the potential has several components each describing a part of the system. Such a dynamical model composed of four components, that is a disk, a nucleus, a bulge and a dark halo was studied by [\citealp{6}]. A new composite mass model describing motion in axially symmetric galaxies with dark matter was recently presented and studied by [\citealp{7}]. Composite axially symmetric galaxy models describing the orbital motion in the Galaxy were also studied by [\citealp{3}]. In these models the gravitational potential is generated by three superposed disks: one representing the gas layer, one the thin disk and one representing the thick disk.

Another interesting class of galaxy models is the self-consistent models. In order to built a self-consistent model, one must take into account Jeans Theorem (see e.g., [\citealp{5}]). According to this theorem, the distribution function $f$ of a steady state galaxy, depends on the integrals of motion, where only isolating integrals are taken into account. The self-consistent problem is one of the most difficult problems in galactic dynamics. Nevertheless, there are several ways of approaching this problem. For instance, Binney \& Tremaine used two different approaches in order to obtain the distribution function. Using the first way, they start from $f$ in order to produce the mass density $\rho$, while in the second they start from $\rho$ heading to $f$.

An effective technique to build self-consistent models for galaxies is the Schwarzschild's orbit superposition method. This method has been applied in order to study the dynamical behavior in both axially symmetric and triaxial galaxies in a number of papers (see e.g., [\citealp{17},\citealp{25},\citealp{28},\citealp{30}]). Self-consistent dynamical models for a disk galaxy with a triaxial halo were constructed by [\citealp{31}], where the starting point was a galaxy in equilibrium composed of an axially symmetric disk-bulge and a halo. Then, applying an artificial acceleration he managed to obtain a system in equilibrium with a triaxial halo.

Realistic triaxial models for galaxies with dark matter haloes were provided in [\citealp{24}]. In their article, the authors extended an earlier method used by [\citealp{23}] to three-dimensional systems by replacing the radial with an ellipsoidal symmetry in the mass density. Triaxial galaxy models were also constructed by [\citealp{2}] and [\citealp{19}].

The main goal of this article is to introduce a new composite mass model in order to use it for the dynamical investigation of galaxies. The model consists of two parts. The main galaxy body and a massive dense nucleus. The model can describe disk and elliptical galaxies as well. Furthermore, the model is designed to describe not only galaxies that are close to axial symmetry but also systems with considerable deviation from axial symmetry. Our target is to investigate the regular and chaotic character of orbits in the new model and connect the degree of chaos with the parameter describing the deviation of the model from axial symmetry.

The paper is organized as follows: In Section \ref{galmod} we present the structure and the properties of our new galactic mass model. In Section \ref{CompMeth} we describe the computational methods we used in order to explore the nature of orbits. In the following Section, we investigate how the parameter describing the deviation from axially symmetry influences the regular or chaotic character of the 3D orbits. We conclude with Section \ref{discus}, where the discussion and the conclusions of this research are presented.

\section{Presentation and analysis of the new dynamical model}
\label{galmod}

Our new dynamical model consists of two components and the total potential $V$ is given by the equation
\begin{equation}
V(x,y,z) = V_G(x,y,z) + V_n(x,y,z),
\label{Vtot}
\end{equation}
where
\begin{equation}
V_{\rm G}(x,y,z) = \frac{- G M_{\rm G}}{\sqrt{b^2 + x^2 + \lambda y^2+ \left(\alpha + \sqrt{h^2 +
z^2}\right)^2}},
\label{Vgal}
\end{equation}
and
\begin{equation}
V_{\rm n}(x,y,z) = \frac{- G M_{\rm n}}{\sqrt{x^2 + y^2 + z^2 + c_{\rm n}^2}}.
\label{Vnuc}
\end{equation}
The first part, Eq. (\ref{Vgal}), is a generalization of the [\citealp{18}] potential (see also [\citealp{8},\citealp{9}]). In Eq. (\ref{Vgal}) $G$ is the gravitation constant, $M_G$ is the mass of the galaxy, while $\alpha, b, h$ and $\lambda$ are parameters connected to the geometry of the galaxy. The parameter $\lambda$ describes the deviation from the axial symmetry. In the case where $b > \alpha$ and $\alpha \gg h$  the model describes a disk galaxy with a disk halo. Here, $\alpha$ is the disk's scale-length, $h$ is the disk's scale-height and $b$ is the core radius of the disk halo. On the other hand, when $b = 0$ and $h \gg \alpha$ the model describes an elliptical galaxy with $\alpha$ and $h$ being the horizontal and vertical scale-lengths respectively. The second part, Eq. (\ref{Vnuc}), is the potential of a spherically symmetric nucleus in which $M_n$ and $c_n$ are the mass and the scale-length of the nucleus respectively. This potential has been used in the past to model the central mass component of a galaxy (see, e.g., [\citealp{14},\citealp{15},\citealp{36}]). Here we must point out, that potential (\ref{Vnuc}) is not intended to represent a black hole nor any other compact object, but a dense and massive nucleus therefore, we don't include any relativistic effects.

We made this choice for two basic reasons. The first reason is that the majority of galaxies are not exactly axially symmetric. The axial symmetry is only a good approximation, allowing us to perform the numerical calculations a lot easier. Therefore, the new model is useful because it is more realistic. Furthermore, this model can be use to describe a wide variety of galaxies. For instance, when $\lambda$ is  close to unity it can describe a nearly axially symmetric galaxy, while for larger values of $\lambda$ it describes systems that are far from axial symmetry. A second reason is that the regular or chaotic nature of orbits is drastically affected by the value of the parameter $\lambda$. Thus, using this new model we can draw useful conclusions connecting the deviation from axial symmetry with the character of orbits in non axially symmetric galaxies.

\begin{figure*}
\resizebox{\hsize}{!}{\includegraphics{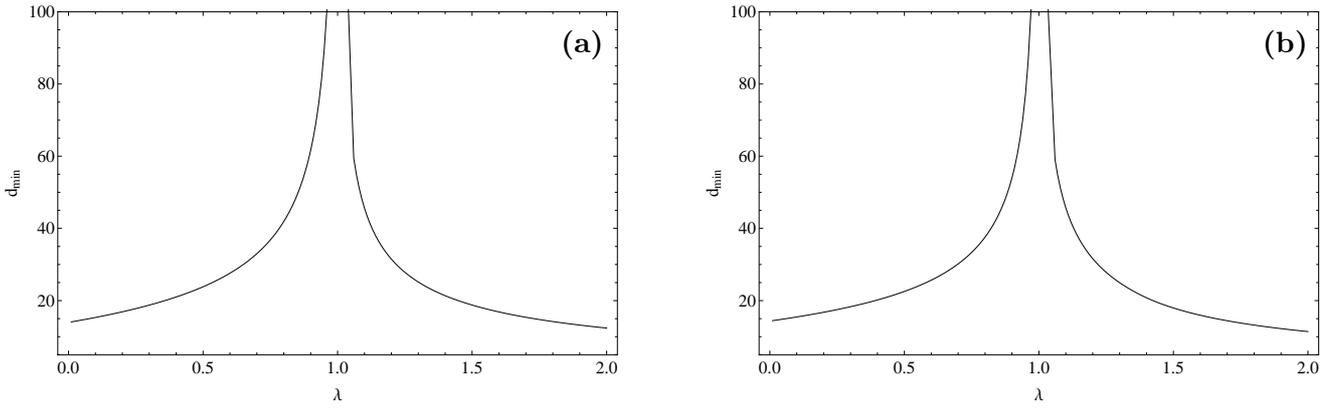}}
\caption{The evolution of the minimum distance $d_{min}$ at which negative value of density appears for the first time, as a function of the parameter $\lambda$ for (a-left): the disk galaxy model (b-right): the elliptical galaxy model.}
\label{deneg}
\end{figure*}

In this work, we use the well known system of galactic units, where the unit of length is 1 kpc, the unit of mass is $2.325 \times 10^7 {\rm M}_\odot$ and the unit of time is $0.9778 \times 10^8$ yr. The velocity units is 10 km/s, while $G$ is equal to unity. The energy unit (per unit mass) is 100 km$^2$s$^{-2}$. In these units the values of the involved parameters are: $M_G = 9500$, $M_n = 400$ and $c_n = 0.25$. For the disk model we choose $b = 12, \alpha = 3$ and $h = 0.15$, while for the elliptical model we have set $b = 0$, $\alpha = 1$ and $h = 10$. The particular values of the system's parameters were chosen having in mind a Milky Way-type galaxy [\citealp{1}]. The parameter describing the deviation from axially symmetry $\lambda$, on the other hand, is treated as a parameter and its value varies in the interval $0.2 \leq \lambda \leq 1.5$.

At this point, we must emphasize, the mass density in our new model (\ref{Vtot}) can be negative when the distance from the centre of the galaxy $d = \sqrt{x^2 + y^2 + z^2}$ described by the model exceeds a minimum distance $d_{min}$, which strongly depends on the parameter $\lambda$. Fig. \ref{deneg}(a-b) shows a plot of $d_{min}$ vs $\lambda$ for the both disk and elliptical galaxy models. In all cases, we consider that the dimensions of our new model do not exceed $d_{min}$. Therefore, the mass density is always positive inside the galaxy described by the model, while is zero elsewhere. Being more precise, our gravitational potential is truncated at $d_{\rm max} = 20$ kpc for both reasons: (i) otherwise the total mass of the galaxy modeled by this potential would be infinite, which is obviously not physical and (ii) to avoid the existence of any negative density.

The corresponding equations of motion are
\begin{equation}
\ddot{x} = - \frac{\partial V}{\partial x}, \ \ \
\ddot{y} = - \frac{\partial V}{\partial y}, \ \ \
\ddot{z} = - \frac{\partial V}{\partial z}.
\label{eqmot}
\end{equation}
The evolution of a deviation vector $\delta \vec{\rm v} = (\delta x, \delta y, \delta z, \delta \dot{x}, \delta \dot{y}, \delta \dot{z})$, which joins the corresponding phase space points of two initially nearby orbits, needed for the computation of the standard indicators of chaos (the SALI in our case), can be monitored by the variational equations
\begin{eqnarray}
\dot{(\delta x)} &=& \delta \dot{x}, \ \ \
\dot{(\delta y)} = \delta \dot{y}, \ \ \
\dot{(\delta z)} = \delta \dot{z}, \nonumber \\
(\dot{\delta \dot{x}}) &=& -\frac{\partial^2 V}{\partial x^2}\delta x - \frac{\partial^2 V}{\partial x \partial y}\delta y -\frac{\partial^2 V}{\partial x \partial z}\delta z, \nonumber \\
(\dot{\delta \dot{y}}) &=& -\frac{\partial^2 V}{\partial y \partial x}\delta x - \frac{\partial^2 V}{\partial y^2}\delta y -\frac{\partial^2 V}{\partial y \partial z}\delta z, \nonumber \\
(\dot{\delta \dot{z}}) &=& -\frac{\partial^2 V}{\partial z \partial x}\delta x - \frac{\partial^2 V}{\partial z \partial y}\delta y -\frac{\partial^2 V}{\partial z^2}\delta z. \nonumber \\
\label{variac}
\end{eqnarray}

The Hamiltonian which determines the motion of a test particle (star) in our system is
\begin{equation}
H = \frac{1}{2}\left(\dot{x}^2 + \dot{y}^2 + \dot{z}^2 \right) + V(x,y,z) = E,
\label{ham}
\end{equation}
where where $\dot{x}$, $\dot{y}$ and $\dot{z}$ are the momenta per unit mass conjugate to $x$, $y$ and $z$ respectively, while $E$ is the numerical value of the Hamiltonian, which is conserved.

\section{Computational techniques}
\label{CompMeth}

When studying the orbital structure of a dynamical system, knowing whether an orbit is regular or chaotic is an issue of significant importance. Over the years, several dynamical indicators have been developed in order to determine the nature of orbits. In our case, we chose to use the Smaller ALingment Index (SALI) method. The SALI [\citealp{27}] is undoubtedly a very fast, reliable and effective tool, which is defined as
\begin{equation}
\rm SALI(t) \equiv min(d_-, d_+),
\label{sali}
\end{equation}
where $d_- \equiv \| {\vec{w_1}}(t) - {\vec{w_2}}(t) \|$ and $d_+ \equiv \| {\vec{w_1}}(t) + {\vec{w_2}}(t) \|$ are the alignments indices, while ${\vec{w_1}}(t)$ and ${\vec{w_2}}(t)$, are two deviations vectors which initially point in two random directions. For distinguishing between ordered and chaotic motion, all we have to do is to compute the SALI for a relatively short time interval of numerical integration $t_{max}$. More precisely, we track simultaneously the time-evolution of the main orbit itself as well as the two deviation vectors ${\vec{w_1}}(t)$ and ${\vec{w_2}}(t)$ in order to compute the SALI. The variational equations (\ref{variac}), as usual, are used for the evolution and computation of the deviation vectors.

\begin{figure}
\includegraphics[width=\hsize]{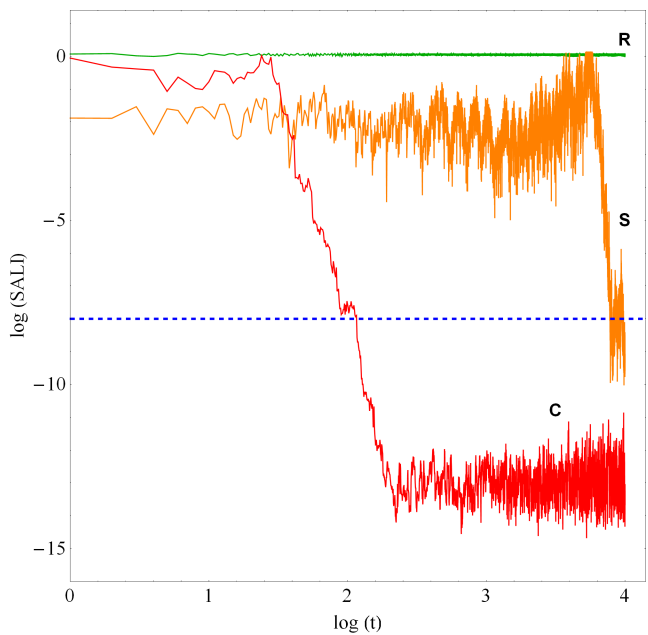}
\caption{Evolution of the SALI of a regular orbit (green color - R), a sticky orbit (orange color - S) and a chaotic orbit (red color - C) in our model for a time period of $10^4$ time units. The horizontal, blue, dashed line corresponds to the threshold value $10^{-8}$ which separates regular from chaotic motion. The chaotic orbits needs only about 120 time units in order to cross the threshold value, while on the other hand, the sticky orbit requires a vast integration time of about 8000 time units so as to reveal its chaotic nature.}
\label{SALIevol}
\end{figure}

The time-evolution of SALI strongly depends on the nature of the computed orbit since when the orbit is regular the SALI exhibits small fluctuations around non zero values, while on the other hand, in the case of chaotic orbits the SALI after a small transient period it tends exponentially to zero approaching the limit of the accuracy of the computer $(10^{-16})$. Therefore, the particular time-evolution of the SALI allow us to distinguish fast and safely between regular and chaotic motion. Nevertheless, we have to define a specific numerical threshold value for determining the transition from regularity to chaos. After conducting extensive numerical experiments, integrating many sets of orbits, we conclude that a safe threshold value for the SALI taking into account the total integration time of $10^4$ time units is the value $10^{-8}$. In order to decide whether an orbit is regular or chaotic, one may use the usual method according to which we check after a certain and predefined time interval of numerical integration, if the value of SALI has become less than the established threshold value. Therefore, if SALI $\leq 10^{-8}$ the orbit is chaotic, while if SALI $ > 10^{-8}$ the orbit is regular. The time evolution of a regular (R) and a chaotic (C) orbit for a time period of $10^4$ time units is presented in Fig. \ref{SALIevol}. The horizontal, blue, dashed line in Fig. \ref{SALIevol} corresponds to that threshold value which separates regular from chaotic motion. Therefore, the distinction between regular and chaotic motion is clear and beyond any doubt when using the SALI method.

In our study, each orbit was integrated numerically for a time interval of $10^4$ time units ($10^{12}$ yr), which corresponds to a time span of the order of hundreds of orbital periods and about 100 Hubble times. The particular choice of the total integration time is an element of great importance, especially in the case of the so called ``sticky orbits" (i.e., chaotic orbits that behave as regular ones during long periods of time). A characteristic example of a sticky orbit (S) in our galactic system can be seen in Fig. \ref{SALIevol}, where we observe that the chaotic character of the particular sticky orbit is revealed after a considerable long integration time of about 8000 time units. A sticky orbit could be easily misclassified as regular by any chaos indicator\footnote{Generally, dynamical methods are broadly split into two types: (i) those based on the evolution of sets of deviation vectors in order to characterize an orbit and (ii) those based on the frequencies of the orbits which extract information about the nature of motion only through the basic orbital elements without the use of deviation vectors.}, if the total integration interval is too small, so that the orbit do not have enough time in order to reveal its true chaotic character. Thus, all the sets of orbits of a given grid were integrated, as we already said, for $10^4$ time units, thus avoiding sticky orbits with a stickiness at least of the order of 100 Hubble times. All the sticky orbits which do not show any signs of chaoticity for $10^4$ time units are counted as regular ones, since that vast sticky periods are completely out of scope of our research.

For the study of our models, we need to define the sample of orbits whose properties (chaos or regularity) we will identify. The best method for this purpose, would have been to choose the sets of initial conditions of the orbits from a distribution function of the models. This, however, is not available so, we define, for each set of values of the parameters of the potential, a grid of initial conditions $(x_0, \dot{x_0})$ regularly distributed in the area allowed by the value of the energy. In each grid the step separation of the initial conditions along the $x$ and $\dot{x}$ axis was controlled in such a way that always there are at least 25000 orbits. For each initial condition, we integrated the equations of motion (\ref{eqmot}) as well as the variational equations (\ref{variac}) using a double precision Bulirsch-Stoer FORTRAN algorithm [\citealp{21}] with a small time step of order of $10^{-2}$, which is sufficient enough for the desired accuracy of our computations (i.e. our results practically do not change by halving the time step). In all cases, the energy integral (Eq. (\ref{ham})) was conserved better than one part in $10^{-10}$, although for most orbits it was better than one part in $10^{-11}$.

\section{Numerical results}
\label{numres}

In this Section, we shall present the results of our research. We start our presentation from the results obtained for the disk galaxy model, which are presented in subsection \ref{diskres}. Moreover, subsection \ref{ellipres} is devoted to the results obtained for elliptical galaxy model. A simple qualitative way for distinguishing between regular and chaotic motion in a Hamiltonian system is by plotting the successive intersections of the orbits using a Poincar\'{e} Surface of Section (PSS) [\citealp{16}]. This method has been extensively applied to 2D models, as in these systems the PSS is a two-dimensional plane. In 3D systems, however, the PSS is four-dimensional and thus the behavior of the orbits cannot be easily visualized.

One way to overcome this issue is to project the PSS to phase spaces with lower dimensions, following the method used in [\citealp{35},\citealp{37}]. Let us start with initial conditions on a 4D grid of the PSS. In this way, we are able to identify again regions of order and chaos, which may be visualized, if we restrict our investigation to a subspace of the whole 6D phase space. We consider orbits with initial conditions $(x_0, z_0, \dot{x_0})$, $y_0 = \dot{z_0} = 0$, while the initial value of $\dot{y_0}$ is always obtained from the energy integral (\ref{ham}). In particular, we define a value of $z_0$, which is kept constant and then we calculate the SALI of the 3D orbits with initial conditions $(x_0, \dot{x_0})$, $y_0 = \dot{z_0} = 0$. Thus, we are able to construct again a 2D plot depicting the $(x, \dot{x})$ plane but with an additional value of $z_0$, since we deal with 3D motion. All the initial conditions of the 3D orbits lie inside the limiting curve defined by
\begin{equation}
f(x,\dot{x};z_0) = \frac{1}{2}\dot{x}^2 + V(x,0,z_0).
\label{zvc}
\end{equation}

For the disk galaxy model we use the energy value $-520$, while for the elliptical galaxy model the energy level is equal to $-540$. We chose for both disk and elliptical galaxy models such energy values which correspond to $x_{\rm max} \simeq 15$ kpc when $z_0 = 1$, where $x_{\rm max}$ is the maximum possible value of the coordinate $x$ on the $(x,\dot{x})$ plane.

\subsection{Disk galaxy model}
\label{diskres}

\begin{figure*}
\centering
\resizebox{0.70\hsize}{!}{\includegraphics{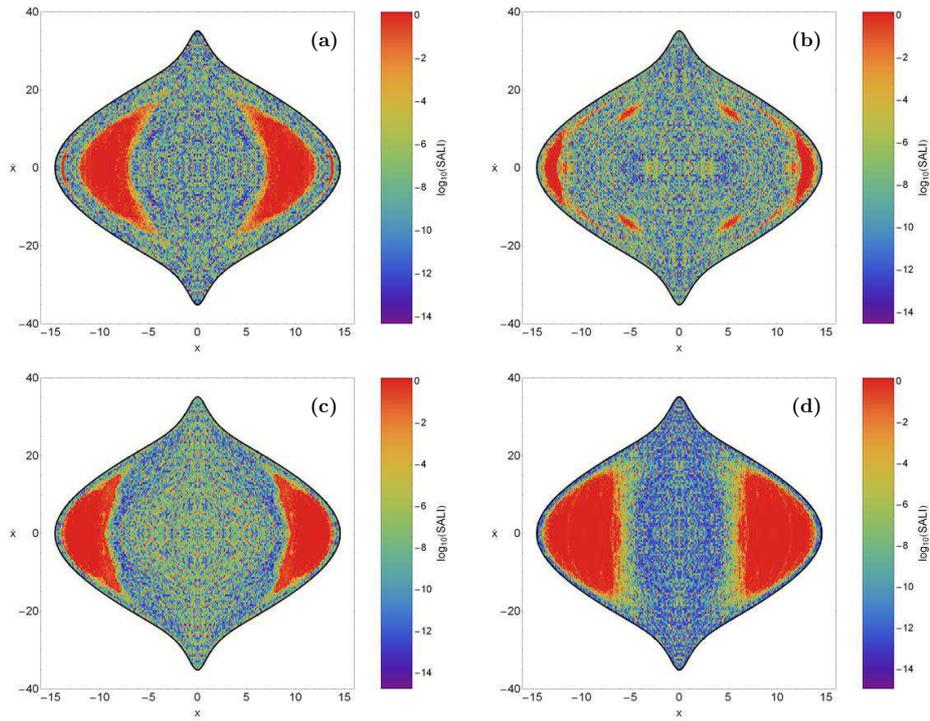}}
\caption{SALI grids of initial conditions $(x_0, \dot{x_0})$ for the disk galaxy model when $z_0 = 1$. (a-upper left): $\lambda = 0.2$, (b-upper right): $\lambda = 0.4$, (c-lower left): $\lambda = 0.6$ and (d-lower right): $\lambda = 0.8$.}
\label{DPz1}
\end{figure*}

\begin{figure*}
\centering
\resizebox{0.70\hsize}{!}{\includegraphics{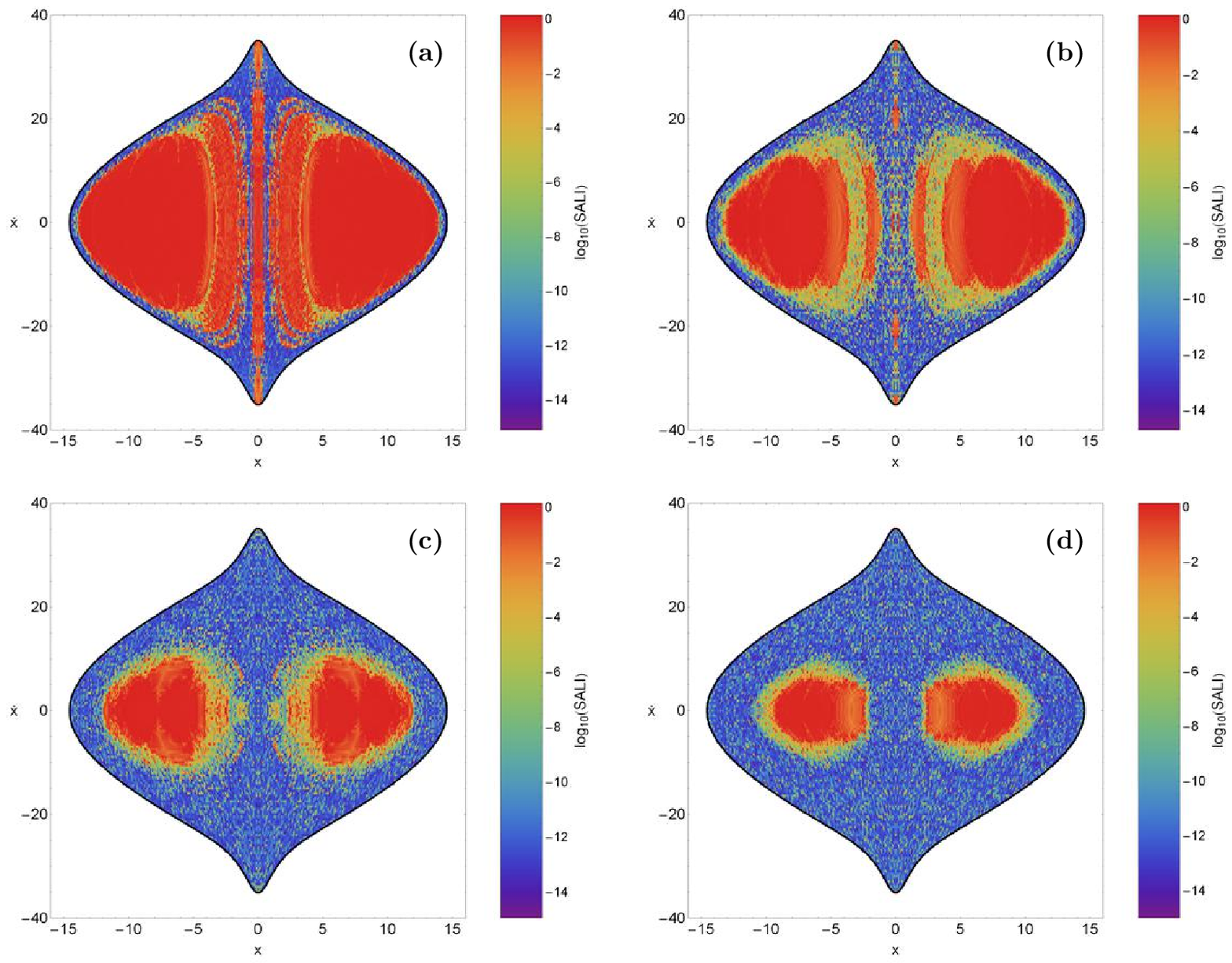}}
\caption{Similar to Fig. \ref{DPz1}(a-d). (a-upper left): $\lambda = 1.01$, (b-upper right): $\lambda = 1.1$, (c-lower left): $\lambda = 1.3$ and (d-lower right): $\lambda = 1.5$.}
\label{DOz1}
\end{figure*}

In Fig. \ref{DPz1}(a-d) we present the final SALI values obtained from the selected grids of initial conditions for four different values of $\lambda$ when $z_0 = 1$. The values of all the other parameters are as in Fig. \ref{deneg}a. Each point is coloured according to its $\rm log_{10}(SALI)$ value at the end of the integration. In these SALI plots, the reddish colors correspond to regular orbits, the blue/purple colors represent the chaotic regions, while all the intermediate colors between the two represent the so-called ``sticky" orbits whose chaotic nature is revealed only after long integration time. The outermost black solid line corresponds to the limiting curve defined by Eq. (\ref{zvc}). Fig. \ref{DPz1}a shows the SALI grid when $\lambda = 0.2$. We observe a vast chaotic sea, which implies that the majority of the orbits are chaotic. However, we can identify several regions of regular motion. The two large islands on either side of the $x = 0$ axis correspond to 2:1 resonant orbits, while the set of the two smaller islands confined to the outer parts of the grid contains initial conditions producing 1:1 resonant orbits. Things are quite different in the grid depicted in Fig. \ref{DPz1}b where $\lambda = 0.4$. Here, the 1:1 resonance is still present, while the 2:1 resonance has disappeared completely. The four regions of stability observed inside the grid correspond to 3:2 resonant 3D orbits. Moreover, proceeding to Fig.\ref{DPz1}c and Fig. \ref{DPz1}d where $\lambda = 0.6$ and $\lambda = 0.8$ respectively, it is evident, that only the 1:1 resonance survives eventually. From the SALI grids presented in Fig. \ref{DPz1}, one may conclude, that for small values of the deviation parameter $\lambda$, our galactic model exhibits several types of resonant orbits, which are strongly affected by shifting the value of $\lambda$. Furthermore, we should point out that as we increase the value of $\lambda$ approaching to axial symmetry $(\lambda = 1)$ the amount of chaos decreases. Similar results can be obtained from the grids shown in Fig. \ref{DOz1}(a-d) where $\lambda = {1.01, 1.1, 1.3, 1.5}$. It is evident, that when $\lambda > 1$ the the 1:1 resonant orbits is the all-dominant type of regular 3D orbits. As the value of $\lambda$ increases (this time moving away from axial symmetry), the chaotic motion grows in percentage at the expense of the 1:1 resonant orbits.

\begin{figure*}
\centering
\resizebox{0.70\hsize}{!}{\includegraphics{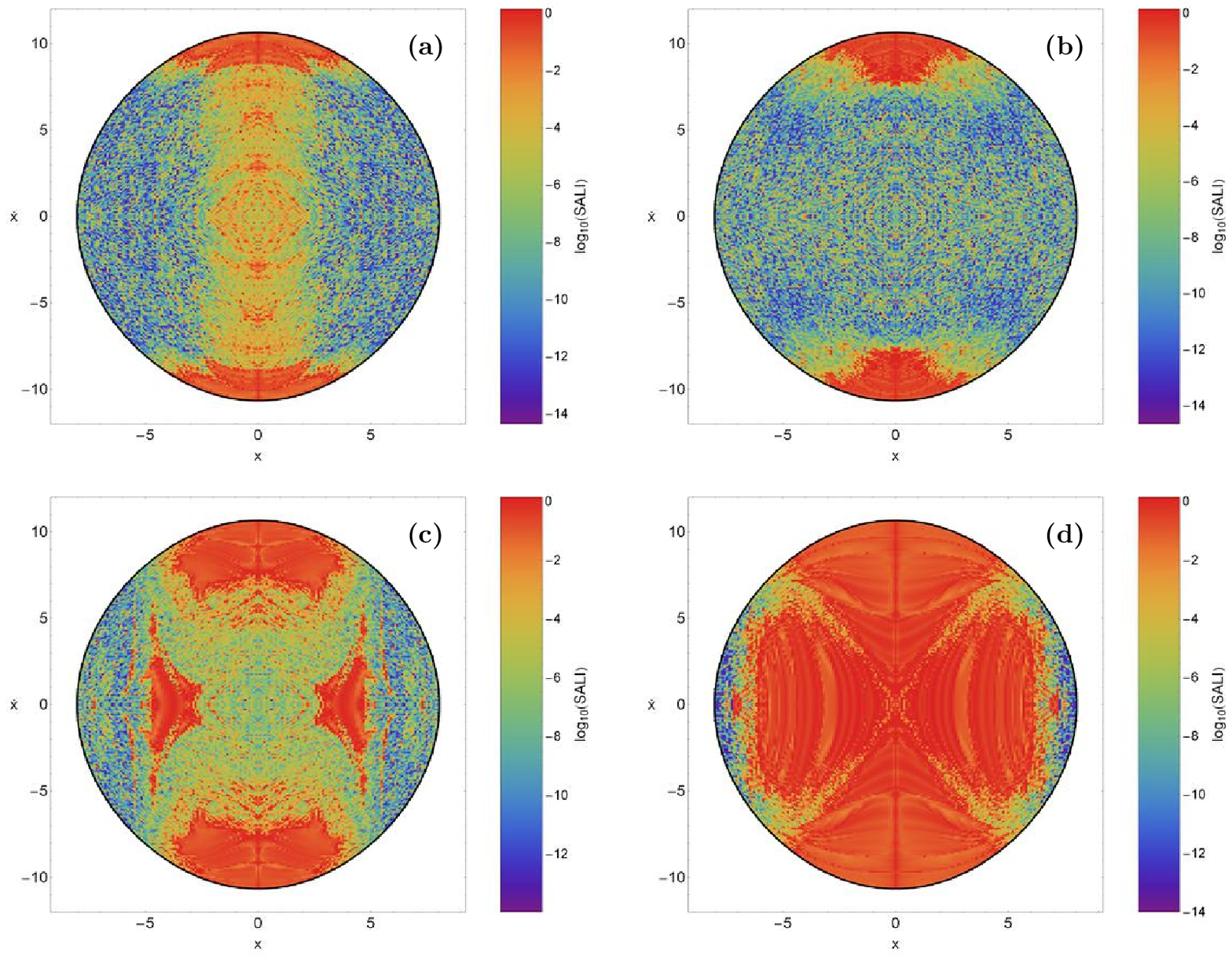}}
\caption{SALI grids of initial conditions $(x_0, \dot{x_0})$ for the disk galaxy model when $z_0 = 10$. (a-upper left): $\lambda = 0.2$, (b-upper right): $\lambda = 0.4$, (c-lower left): $\lambda = 0.6$ and (d-lower right): $\lambda = 0.8$.}
\label{DPz10}
\end{figure*}

\begin{figure*}
\centering
\resizebox{0.70\hsize}{!}{\includegraphics{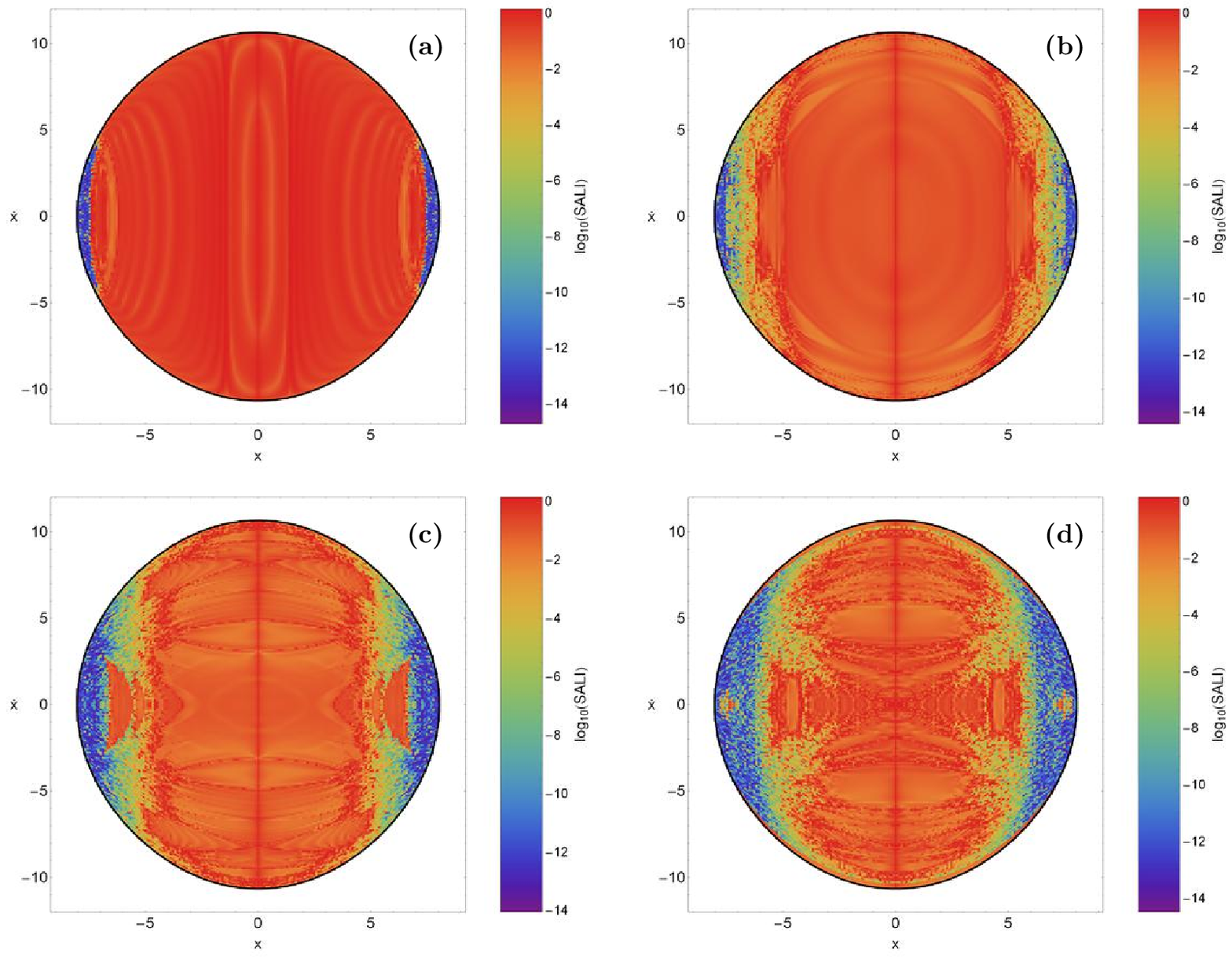}}
\caption{Similar to Fig. \ref{DPz10}(a-d). (a-upper left): $\lambda = 1.01$, (b-upper right): $\lambda = 1.1$, (c-lower left): $\lambda = 1.3$ and (d-lower right): $\lambda = 1.5$.}
\label{DOz10}
\end{figure*}

\begin{figure*}
\centering
\resizebox{0.90\hsize}{!}{\includegraphics{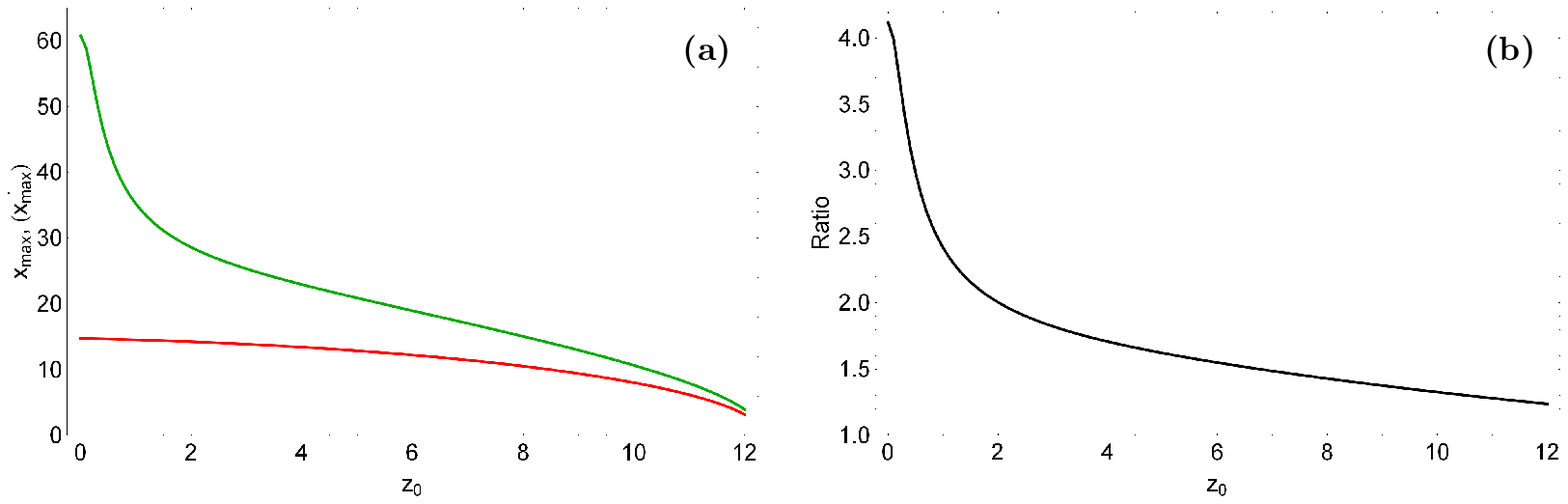}}
\caption{(a-left): Evolution of the maximum value of the $x$ coordinate $x_{max}$ (red color) and the maximum value of the $\dot{x}$ velocity $\dot{(x_{max})}$ (green color) on the $(x,\dot{x})$ plane versus $z_0$ and (b-right): correlation between the ratio $\dot{(x_{max})}/x_{max}$ and the $z_0$ value in the $\lambda = 1.5$ disk galaxy model.}
\label{theors}
\end{figure*}

We proceed now, investigating how the deviation parameter $\lambda$ influences 3D orbits with large values of $z_0$. We choose $z_0 = 10$ as a fiducial value for our sets of initial conditions of orbits. Fig. \ref{DPz10}(a-d) shows the SALI grids of initial conditions when $\lambda = {0.2, 0.4, 0.6, 0.8}$. In Fig. \ref{DPz10}a where $\lambda = 0.2$ we see that the chaotic orbits are abundant producing a large chaotic sea, while at the central parts of the grid there is a substantial amount of sticky orbits corresponding to intermediate colors of SALI. On the other hand, the ordered motion is confined mainly at the outer upper and lower parts of the grid. When $\lambda = 0.4$ we observe in Fig. \ref{DPz10}b, that the percentage of the sticky orbits has been decreased drastically and therefore, the structure of the chaotic domain is much more solid and well defined. Things however, become very interesting in Fig. \ref{DPz10}c where $\lambda = 0.6$. Here, sticky orbits grow again in percentage and also two additional regions of regular motion emerge. In Fig. \ref{DPz10}d where $\lambda = 0.8$ we observe, that the initial conditions corresponding to regular motion have inundated almost the entire grid, thus confining chaotic orbits to the right and left outer regions of the plane. It is worth noticing, that when $z_0 = 10$ the available area inside the $(x,\dot{x})$ plane is considerably less than that when $z_0 = 1$. What actually happens is that by shifting to higher levels of $z_0$, both the maximum possible value of the $x$ coordinate and the maximum possible value of the $\dot{x}$ velocity on the $(x,\dot{x})$ plane are constantly reduced as it can be seen in Fig. \ref{theors}a where a plot of $x_{max}$ (red color) and $\dot{(x_{max})}$ (green color) versus $z_0$ is presented. Moreover, for large values of $z_0$ we see that the outermost limiting curve tends to be circular, because the ratio $\dot{(x_{max})}/x_{max}$ tends asymptotically to unity with increasing $z_0$ (see Fig. \ref{theors}b). Similar behavior applies in the case of the elliptical galaxy. Thus, we could argue, that this fact justifies in a way, the swarming percentages of regular orbits (see e.g., Fig. \ref{DPz10}d). Quite similar outcomes are obtained when $\lambda > 1$ and $z_0 = 10$. Fig. \ref{DOz10}a shows the grid when $\lambda = 1.01$, that is an almost axially symmetric model. As expected, the vast majority of the computed orbits are regular, while the small deviation from axially symmetry induces a low percentage of chaos, which is confined at the left and right regions of the grid. Moreover, as the value of the deviation parameter $\lambda$ increases, there is a constant increase of the portion of the initial conditions of chaotic orbits, which penetrates deeper and deeper toward the central region of the grid (see, Figs. \ref{DOz10}(b-d)).

\begin{figure}
\includegraphics[width=\hsize]{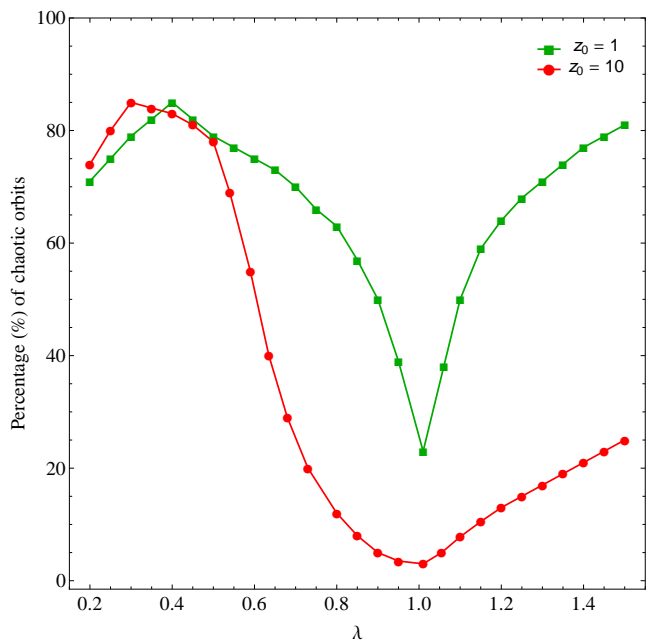}
\caption{Evolution of the percentage of chaotic orbits in the disk galaxy models as a function of the deviation parameter $\lambda$. Green color corresponds to $z_0 = 1$, while red color to $z_0 = 10$.}
\label{disk_ch}
\end{figure}

Thus, calculating from all the SALI grids the percentage of chaotic orbits (including also the sticky orbits, which if we integrate them using more time will eventually reveal their chaotic nature), we are able to follow how this fraction varies as a function of the deviation parameter $\lambda$. As it is clear from Fig. \ref{disk_ch}, the pattern of the evolution of the chaotic percentage is quite similar in both studied cases ($z_0 = 1$ and $z_0 = 10$). In particular, wee see, that the amount of chaos decreases sharply when $0.35 \lesssim \lambda < 1$, while this tendency is reversed when $\lambda > 1$. For $\lambda > 0.4$ the chaotic percentage held always larger values when $z_0 = 1$ than when $z_0 = 10$. Clearly, the minimum value of the chaotic percentage is observed when the galactic model is axially symmetric $(\lambda = 1)$. Specifically, in the case where $z_0 = 10$, the chaotic percentage tends asymptotically to zero when $\lambda = 1$. On the contrary, chaos remains at high levels around 25\% when $z_0 = 1$, probably due to the greater influence of the massive nucleus.

\subsection{Elliptical galaxy model}
\label{ellipres}

\begin{figure*}
\centering
\resizebox{0.70\hsize}{!}{\includegraphics{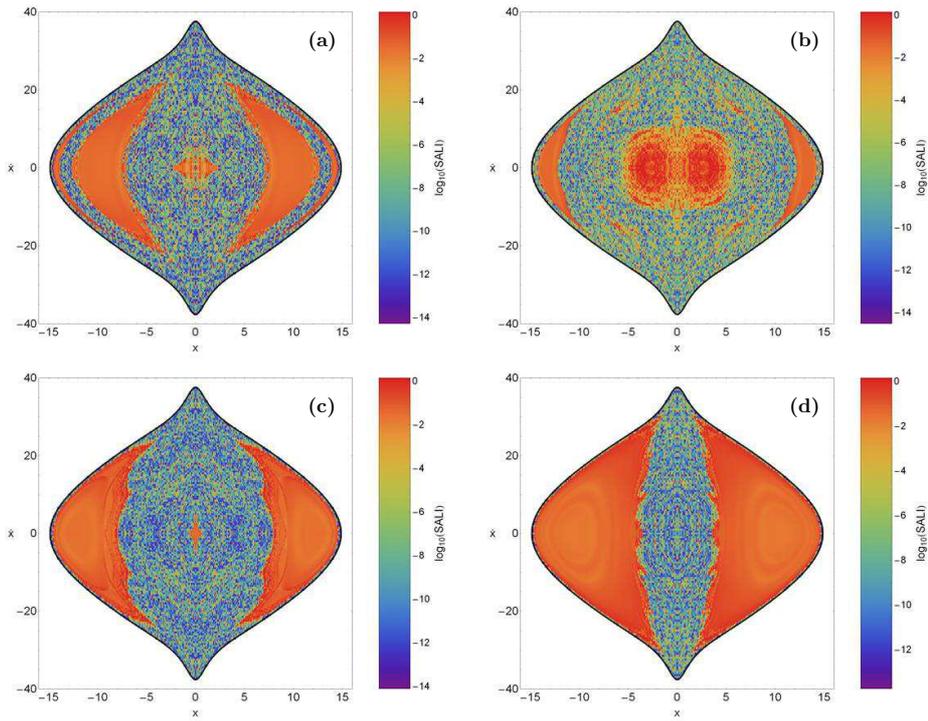}}
\caption{SALI grids of initial conditions $(x_0, \dot{x_0})$ for the elliptical galaxy model when $z_0 = 1$. (a-upper left): $\lambda = 0.2$, (b-upper right): $\lambda = 0.4$, (c-lower left): $\lambda = 0.6$ and (d-lower right): $\lambda = 0.8$.}
\label{EPz1}
\end{figure*}

We now turn our investigation to the elliptical galaxy model. Once more, we define sets of initial conditions and then we compute the SALI of the 3D orbits in an attempt to construct grids thus distinguishing between regular and chaotic motion. In Fig. \ref{EPz1}(a-d) we present such SALI grids of initial conditions for four different values of $\lambda$ when $z_0 = 1$. The values of all the other parameters are as in Fig. \ref{deneg}b corresponding to the elliptical galaxy model. Fig. \ref{EPz1}a depicts the structure of the $(x, \dot{x})$ plane when $\lambda = 0.2$. We see that a large unified chaotic sea exists which however, surrounds several regions of regular motion. We identify three different types of regular 3D orbits: (i) 1:1 resonant orbits producing the set of the two islands of stability located at the outer parts of the grid, (ii) 2:1 resonant orbits with initial conditions inside the two big and distinct islands and (iii) 3:1 resonant orbits, which produce the two small islands near the center. Things are quite different though according to Fig. \ref{EPz1}b where $\lambda = 0.4$. Here, the islands of the 1:1 resonance at the outer parts of the grid have been increased in size, while the 2:1 resonance occupy now the central region replacing, in a way, the 3:1 resonance. With a more closer look, one may also identify a set of four small islands of stability above the 2:1 resonance. These islands correspond to the 3:2 resonance. In Fig. \ref{EPz1}c where $\lambda = 0.6$ we observe that the percentage of the 1:1 resonant orbits has been increased significantly, while the central region of the 2:1 resonant orbits has been shrank considerably. Therefore, we may conclude, that the increase of the values of $\lambda$ in the elliptical model causes the extinction of secondary resonances, while at the same time reveals the predominance of the 1:1 orbits. Indeed, when $\lambda = 0.8$ we see at Fig. \ref{EPz1}d that the chaotic domain has been confined mainly at the central parts of the grid, while all the secondary resonances are absent due to the growth of the percentage of the 1:1 orbits.

\begin{figure*}
\centering
\resizebox{0.70\hsize}{!}{\includegraphics{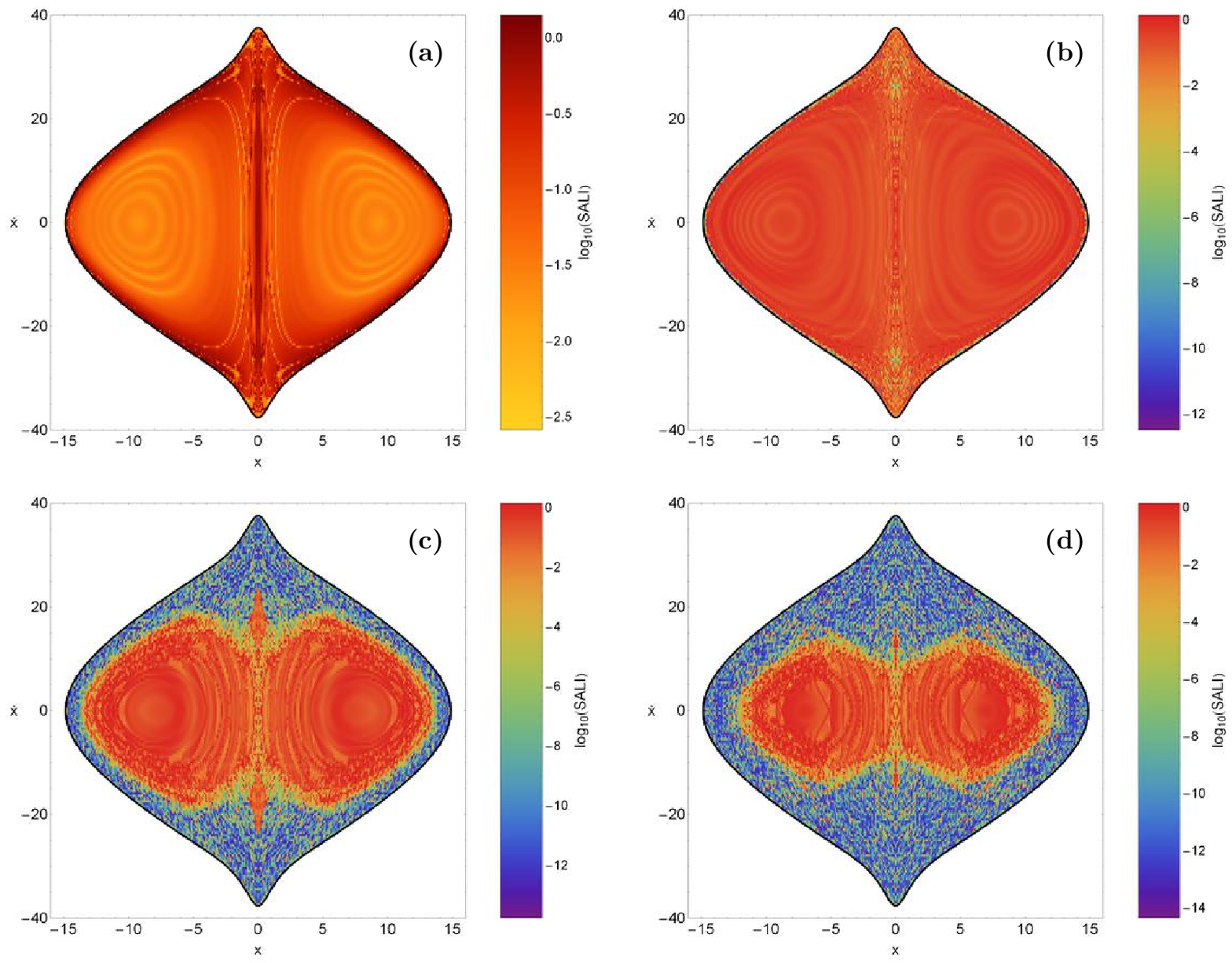}}
\caption{Similar to Fig. \ref{EPz1}(a-d). (a-upper left): $\lambda = 1.01$, (b-upper right): $\lambda = 1.1$, (c-lower left): $\lambda = 1.3$ and (d-lower right): $\lambda = 1.5$.}
\label{EOz1}
\end{figure*}

\begin{figure*}
\centering
\resizebox{0.70\hsize}{!}{\includegraphics{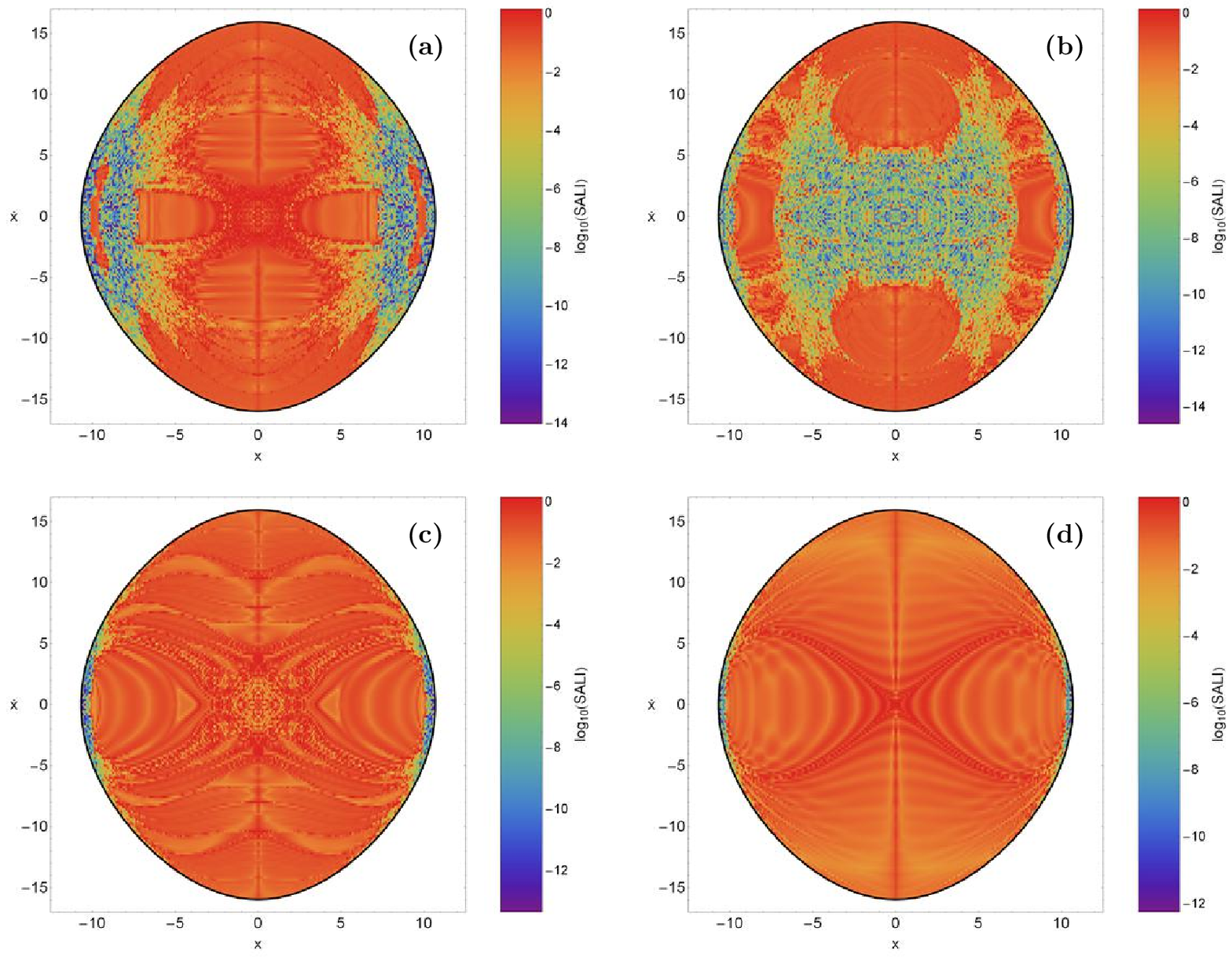}}
\caption{SALI grids of initial conditions $(x_0, \dot{x_0})$ for the elliptical galaxy model when $z_0 = 10$. (a-upper left): $\lambda = 0.2$, (b-upper right): $\lambda = 0.4$, (c-lower left): $\lambda = 0.6$ and (d-lower right): $\lambda = 0.8$.}
\label{EPz10}
\end{figure*}

\begin{figure*}
\centering
\resizebox{0.70\hsize}{!}{\includegraphics{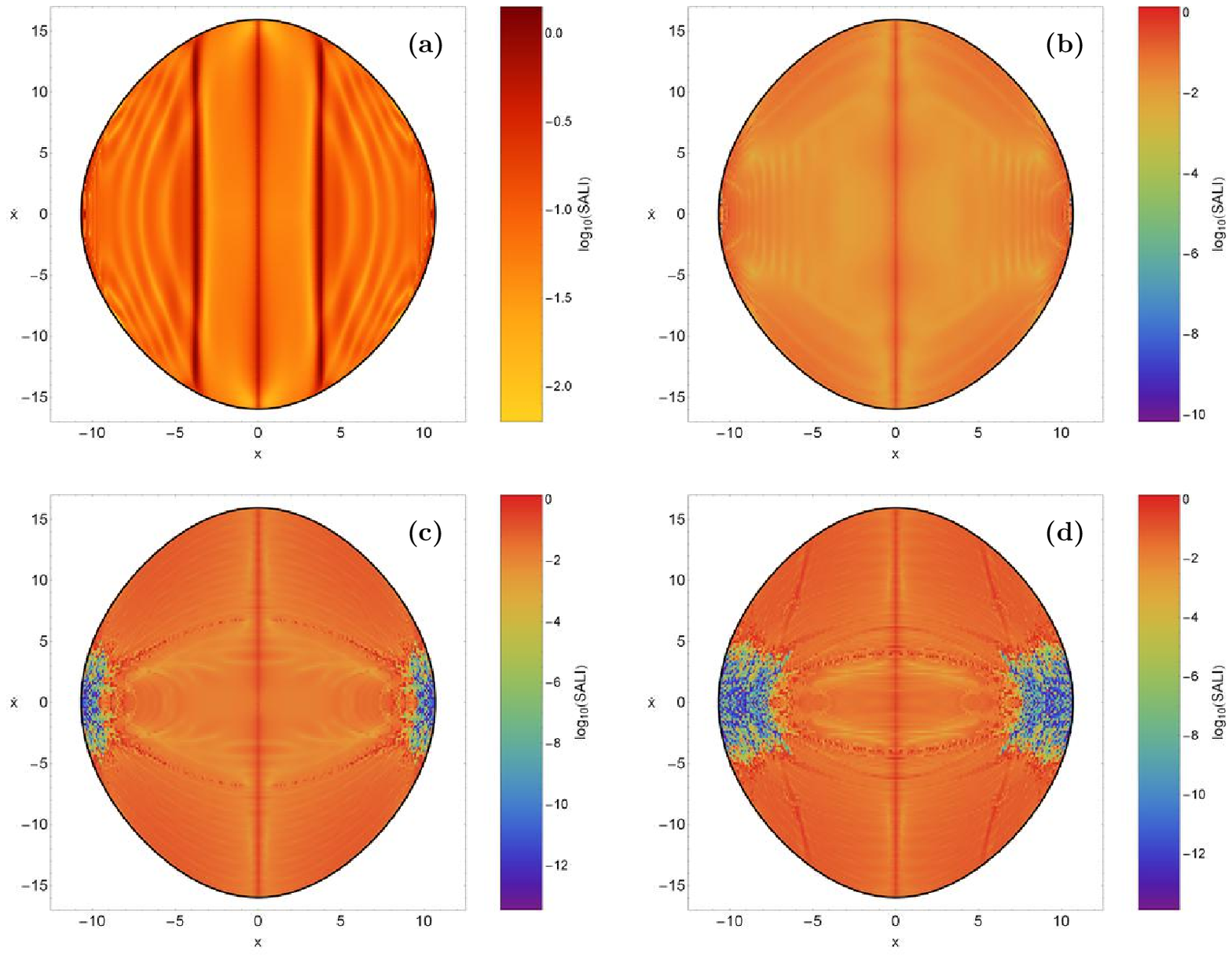}}
\caption{Similar to Fig. \ref{EPz10}(a-d). (a-upper left): $\lambda = 1.01$, (b-upper right): $\lambda = 1.1$, (c-lower left): $\lambda = 1.3$ and (d-lower right): $\lambda = 1.5$.}
\label{EOz10}
\end{figure*}

We continue our investigation presenting in Fig. \ref{EOz1}(a-b) four additional SALI grids when $\lambda = {1.01, 1.1, 1.3, 1.5}$. In Fig. \ref{EOz1}a we see, that even though the model is not axially symmetric, since $\lambda = 1.01$ however, the grid is covered entirely by initial conditions corresponding to 1:1 regular orbits. Furthermore, when $\lambda = 1.1$ we may identify only some nuggets of chaos in Fig. \ref{EOz1}b. Only when $\lambda = 1.3$ a well define and unified chaotic sea establishes in the grid, as we can see in Fig. \ref{EOz1}c. The extent of the chaotic sea increases as we amplify the value of $\lambda$ thus having elliptical galaxy models away from axial symmetry. Taking into account the numerical results presented in Fig. \ref{EOz1}(a-b) when $\lambda > 1$ we may draw the following conclusions: (i) there is no evidence of secondary resonances, (ii) the 1:1 resonance is the dominant, if not the only, type of regular motion and (iii) the increase of the value of the deviation parameter has as results the increase of the amount of chaos.

Our next step, is to determine how the deviation parameter $\lambda$ influences the regular or chaotic character of the 3D orbits when $z_0 = 10$. In Fig. \ref{EPz10}(a-d) we provide four SALI grids corresponding to $\lambda = 0.2, 0.4, 0.6, 0.8$ respectively. When $\lambda = 0.2$ we see that the vast majority of the grid is covered by initial conditions producing regular orbits, while the chaotic initial conditions are confined at the left and right outer parts of the grid. Surprisingly enough, we observe in Fig. \ref{EPz10}b where $\lambda = 0.4$ that all the central initial conditions have now changed their nature from regular to chaotic. This interplay continues in reverse order in Fig. \ref{EPz10}c where $\lambda = 0.6$. Here, the same orbits with initial conditions located at the central region of the grid have changed once more their character this time from chaotic to regular. Moreover, all the chaotic initial conditions are confined at the outer parts of the grid. Things are quite similar when $\lambda = 0.8$. Thus, we may conclude that for small values of $\lambda$ there is an interplay between regular and chaotic motion, while for larger values tending to axial symmetry, almost all the orbits are 1:1 regular orbits. In Fig. \ref{EOz10}(a-b) we present the structure of SALI grids when $\lambda > 1$. We see, that in fact, the results are very similar to those discussed earlier in Fig. \ref{EOz1}(a-d). In particular, when $\lambda = 1.01$ the entire grid is occupied only by initial conditions corresponding to regular orbits. However, the amount of chaos is increasing slowly but steadily as we proceed to models with larger values of deviation parameter (see Fig. \ref{EOz10}(b-d)).

\begin{figure}
\includegraphics[width=\hsize]{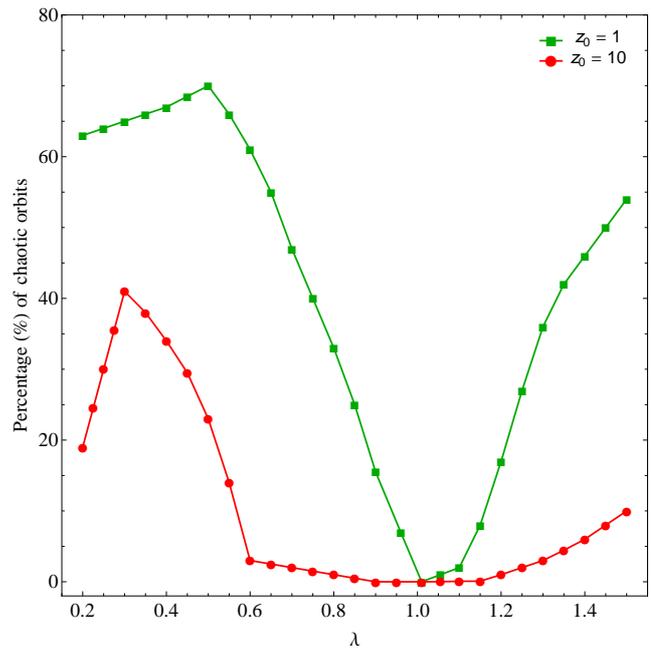}
\caption{Evolution of the percentage of chaotic orbits in the elliptical galaxy models as a function of the deviation parameter $\lambda$. Green color corresponds to $z_0 = 1$, while red color to $z_0 = 10$.}
\label{ellip_ch}
\end{figure}

\begin{figure*}
\centering
\resizebox{0.90\hsize}{!}{\includegraphics{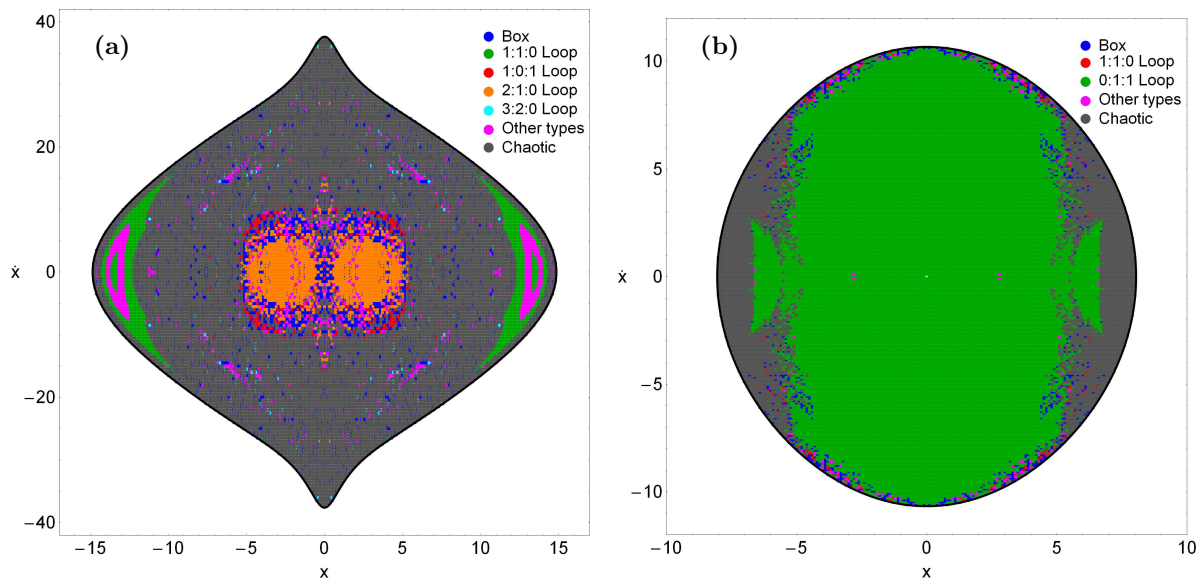}}
\caption{Orbital structure of the $(x,\dot{x})$ plane for (a-left): an elliptical galaxy model with $\lambda = 0.4$ and $z_0 = 1$; (b-right): a disk galaxy model with $\lambda = 1.3$ and $z_0 = 10$.}
\label{clas}
\end{figure*}

\begin{figure*}
\centering
\resizebox{0.85\hsize}{!}{\includegraphics{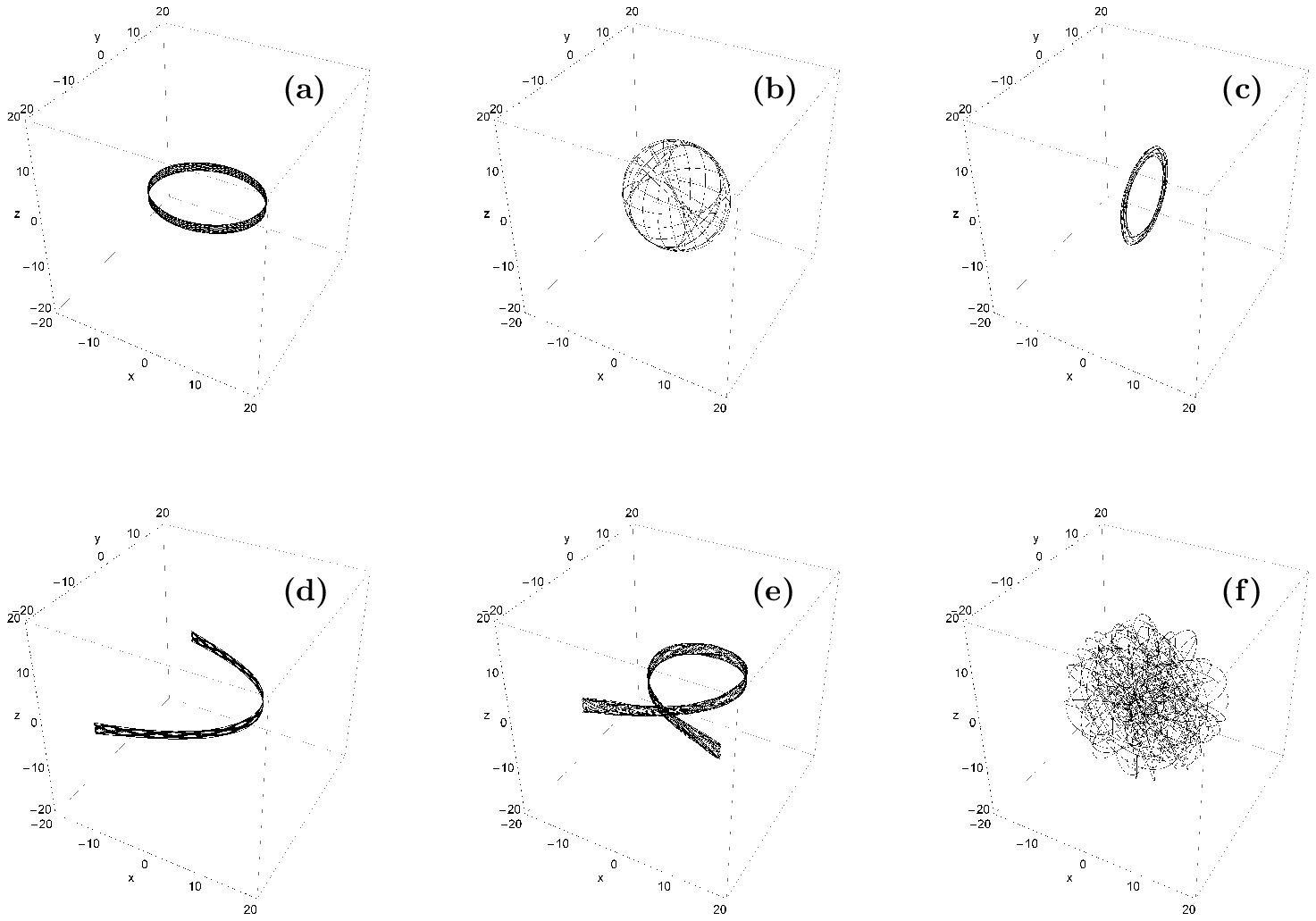}}
\caption{Characteristic examples of 3D orbits encountered in our disk/elliptical galaxy models. More details are given in the text.}
\label{orbs}
\end{figure*}

Of particular interest is the evolution of the percentage of chaotic orbits in the elliptical galaxy models as a function of the deviation parameter $\lambda$, which is presented in Fig. \ref{ellip_ch}. We observe, that the basic pattern of the evolution is the same in both cases ($z_0 = 1$ and $z_0 = 10$). In fact, we may distinguish three different regions at the curves: (i) for $0.2 < \lambda \lesssim 0.4$ the chaotic percentage increases; (ii) for $\lambda > 0.4$ the amount of chaos exhibits a rapid decrease until $\lambda = 1$ where chaos vanishes due to the axial symmetry of the system and the two curves coincide; (iii) for $\lambda > 1$ the percentage of chaotic orbits increases following a monotonic trend. Here, we should point out that when $z_0 = 1$ the amount of chaos is larger, while the changes to the chaos evolution (increase of decrease) are more sharp than when $z_0 = 10$.

\begin{table}
\begin{center}
   \caption{Type, model and initial conditions for the 3D orbits shown in Fig. \ref{orbs}(a-f). In all cases, $y_0 = \dot{z_0} = 0$, while $\dot{y_0}$ is found from the energy integral given by Eq. (\ref{ham}).}
   \label{table}
   \begin{tabular}{@{}lcccccc}
      \hline
      Figure & Type of orbit & model & $\lambda$ & $x_0$ & $z_0$ & $\dot{x_0}$  \\
      \hline
      \ref{orbs}a & 1:1:0 & disk & 0.8 & 10.6 & 1 & 0.00 \\
      \ref{orbs}b & 1:0:1 & elliptical & 0.8 & 0.00 & 10 & 10.2 \\
      \ref{orbs}c & 0:1:1 & disk & 1.3 & 0.10 & 10 & 0.00 \\
      \ref{orbs}d & 1:2:0 & disk & 0.2 & 9.85 & 1 & 0.00 \\
      \ref{orbs}e & 2:3:0 & disk & 0.4 & 5.10 & 1 & 14.4 \\
      \ref{orbs}f & chaotic & elliptical & 0.6 & 1.50 & 10 & 0.00 \\
      \hline
   \end{tabular}
\end{center}
\end{table}

In all cases (disk and elliptical galaxy models), we further classified the regular orbits into different families, by using the technique of frequency analysis used by [\citealp{10},\citealp{20}]. Initially, [\citealp{4}] proposed a technique, dubbed spectral dynamics, for this particular purpose. Later on, this method has been extended and improved by [\citealp{10}], while the extraction of basic frequencies was obtained with the frequency modified Fourier transform which was refined by [\citealp{26}]. In general terms, this method computes the Fourier transform of the coordinates of an orbit, identifies its peaks, extracts the corresponding frequencies and search for the fundamental frequencies and their possible resonances. Thus, we can easily identify the various families of regular orbits and also recognize the secondary resonances that bifurcate from them. In Fig. \ref{clas}(a-b) we present two characteristic examples thus demonstrating the reconstruction of the orbital structure of the $(x,\dot{x})$ plane, which enable us now to distinguish not only between regular and chaotic motion but also between different families of regular orbits. Fig. \ref{clas}a depicts the case of an elliptical galaxy model with $\lambda = 0.4$ and $z_0 = 1$, while in Fig. \ref{clas}b we see a similar grid for a disk galaxy model with $\lambda = 1.3$ and $z_0 = 10$. Here we should point out, that these two grids are in fact advanced versions of the regular/chaotic SALI grids given in Fig. \ref{EOz1}c and Fig. \ref{DPz10}b respectively.

We shall close this section by presenting in Fig. \ref{orbs}(a-f) several characteristic examples of 3D orbits that are encountered in our galaxy model. All orbits were computed until $t = 200$ time units. The exact type (resonance), the model used to produce each orbits and the initial conditions are given in Table \ref{table}. It is worth noticing that the 1:1 resonance is usually the hallmark of loop orbits, both coordinates oscillating with the same frequency in their main motion. Previously, we have seen that in general terms, the dominant type of regular orbits is the 1:1 resonant orbits. Our extensive numerical calculations indicate, that the 1:1 resonance in our model appears in three different forms, according to which axes the oscillations take place. In particular, we see in Figs. \ref{orbs}(a-c) three different types of 1:1 resonant orbits. In Fig. \ref{orbs}a we have the 1:1:0 resonance since the oscillations take place to the $x$ and $y$ axes. On the other hand, in Fig. \ref{orbs}b we see a 1:0:1 resonant orbit oscillating at $x$ and $z$ axes, while in Fig. \ref{orbs}c the oscillations take place at the $y$ and $z$ axis and therefore we have a 0:1:1 resonant 3D orbit. The first type, that is the 1:1:0, is very common to both disk and elliptical galaxy models with low values of $z$. In galaxy models with large values of $z_0$ (i.e., $z_0 = 10$) the dominant resonance transforms to the other two types (1:0:1 and 0:1:1). So, one may conclude, that orbits possessing low values of $z_0$ should circulate horizontally (parallel with the galactic plane) around the nucleus, while for 3D orbits having large values of $z_0$ we expect them to perform loop orbits perpendicularly to the galactic plane.

\section{Discussion}
\label{discus}

Astronomers built and use dynamical models in order to represent and therefore study the structure and the evolution of galaxies. The data needed for the construction of these models, consist mainly of images and spectra obtained using ground-based observations as well as the Hubble Space Telescope (HST).

Axially symmetric models for the central parts of galaxies, containing a central black hole (BH) were constructed by [\citealp{13}]. In their paper, the authors combined ground based data from the Michigan-Dartmouth-MIT (MDM) observatory with similar input from HST. In particular, the technique of numerical orbit superposition was applied, in order to built galactic models with distribution functions with three isolating integrals of motion. Then, the mass of the central BH, the mass to light ratio and also the orbital structure of the system could be obtained from those models.

Moreover, axially symmetric models, for oblate elliptical galaxies, with a distribution function depending on two integrals of motion were constructed by [\citealp{29}]. In their work, the authors used high quality data from the HST. Using the models, they managed to obtain the dynamical mass to light ratio $M/L$ and the corresponding rotation rate of each galaxy. They also found that the brightest galaxies rotate too slow to account for their flattening.

From all the above, it becomes clear that nearly axially symmetric, triaxial and asymmetric galaxies cannot be represented by axially symmetric models. Therefore, it seems necessary to construct a new dynamical model in order to be able to describe the properties of motion in non axially symmetric galaxies.

In this paper, we have presented a new dynamical mass model for non axially symmetric galaxies. The model consists of two parts. The first part represents the main body of the galaxy, while the second part represents a massive and dense central nucleus. We made this choice for a number of reasons. A first reason is that in most galaxies, the axial symmetry is just an approximation in order to make the mathematical study of galaxies more convenient. On the other hand, there is no doubt that there are galaxies that are close to axial symmetry, as well as galaxies that are not axially symmetric. Since our model covers a large variety of types of galaxies, the model could be considered more realistic. For a second reason, we can argue the following: As there is evidence that most galaxies host massive objects [\citealp{32}], in their centres, we constructed a model with a spherical massive nucleus. We believe that with this additional massive nucleus, we describe more precisely a real galaxy. A third reason is that the dense and massive nucleus plays a vital role on the nature of motion, that is the regular or chaotic character of orbits (see [\citealp{6},\citealp{36}] and references therein).

An additional advantage of the new dynamical model is that it can describe motion in disk as well as in elliptical galaxies. This is obtained by suitably choosing the values of the parameters $(\alpha, b, h)$, while the deviation from axial symmetry is regulated by the quantity $\lambda$. Here we must make clear, that we consider that the dimensions of the new galaxy dynamical model are taken such as the mass density is always positive inside the galaxy and zero elsewhere (see Fig. \ref{deneg}a-b).

We would also like to remind to the reader, that we have constructed this model in order to investigate the regular or chaotic nature of orbits and to try to connect it with the parameter $\lambda$. In order to obtain this, we used the SALI method. Using the same technique, we obtained interesting results on describing the different families of regular orbits that are present in the model. Our results referring to the dynamical properties of the new galactic model can be summarized as follows:

\textbf{(1).} Taking into account that our dynamical model is three-dimensional (3D), we had to find a way to define the sample of orbits whose properties (order or chaos) would be examined. A very convenient technique, which is described in Sec. \ref{numres}, was used and thus, we were able to study 3D orbits with initial condition in $(x,\dot{x})$ plane with an additional value of $z_0$. We then studied how the particular value of $z_0$ controls the amount of chaos in the system by choosing two different values of $z_0$, leading to the conclusion that $z_0$ is a key element regarding the nature of orbits.

\textbf{(2).} We conducted a thorough investigation in several cases, using different values of the parameter in the range $0.2 < \lambda < 1.5$. Our numerical results, indicate that the parameter $\lambda$, which describes the deviation from axially symmetry is indeed very influential both in the disk and the elliptical galaxy models.

\textbf{(3).} When $\lambda < 1$ and $z_0 = 1$ we found that in both disk and elliptical galaxy models a lot of different types of resonances appear. On the other hand, in galaxy models with large values of $z_0$, such as $z_0 = 10$, all the secondary resonances are suppressed and the 1:1 loop orbits is the dominant type.

\textbf{(4).} In the case where $\lambda > 1$ the percentage of secondary resonances is extremely low and the 3D 1:1 loop orbits are abundant both in elliptical and disk galaxy models for all values of $z_0$. However, the exact structure of these orbits differ significantly according to the value of $z_0$. In fact, stars moving in 3D regular orbits with low values of $z_0$ circulate parallel to the galactic plane, while for large values of $z_0$ the 3D loop orbits are performed vertically to the galactic plane.

\textbf{(5).} We found a very strong correlation between the value of the deviation parameter $\lambda$ and the fraction of chaotic orbits both in disk and elliptical models. According to our numerical experiments, chaos turns out to be dominant in galaxy models with sufficient deviation from axial symmetry. Moreover, the observed amount of chaos in the disk galaxy models is significantly larger than in elliptical models.

\section*{Acknowledgments}

The authors would like to thank the two anonymous referees for the careful reading of the manuscript, their positive comments and for all the aptly suggestions which allowed us to improve both the quality and the clarity of our work.

\end{document}